\documentclass[aps,prl,twocolumn,showpacs,superscriptaddress]{revtex4-2}

\usepackage{graphicx}
\usepackage{subfigure}
\usepackage{bm}
\usepackage{latexsym,amsthm,amssymb}
\usepackage{booktabs,threeparttable}
\usepackage{makeidx}
\usepackage{amsmath}
\usepackage{amsfonts}
\usepackage{amssymb}
\usepackage{setspace}
\usepackage{slashed}
\usepackage{tikz}
\usepackage{hyperref}
\hypersetup{
    colorlinks=true,
    linkcolor=blue,
    filecolor=magenta,
    urlcolor=cyan,
}

\usepackage{xcolor}
\usepackage{xspace}

\let\originalleft\left
\let\originalright\right
\renewcommand{\left}{\mathopen{}\mathclose\bgroup\originalleft}
\renewcommand{\right}{\aftergroup\egroup\originalright}

\newcommand{\zc}{z_{\rm cut}}

\newcommand{\la}{\lambda_\alpha}

\newcommand{\pythia}{P\protect\scalebox{0.8}{YTHIA}\xspace}

\newcommand{\fastjet}{F\protect\scalebox{0.8}{AST}J\protect\scalebox{0.8}{ET}\xspace}

\newcommand{\softdrop}{\textsf{SoftDrop }}
\newcommand{\fjcontrib}{\textsf{fjcontrib}}

\def\beq{\begin{equation}}  
\def\eeq{\end{equation}}
\def\({\left(}
\def\){\right)}
\def\[{\left[}
\def\]{\right]}

\begin{document}
\begin{minipage}[t]{0.45\textwidth}
    \raggedleft
	JLAB-THY-24-4245
\end{minipage}

\title{Jet substructure of light and heavy flavor jets at RHIC}

\author{Zhuoheng~Yang}
\email{zhuoheng.yang@stonybrook.edu}
\affiliation{Department of Physics and Astronomy, Stony Brook University, Stony Brook, NY 11794, USA.}

\author{Oleh~Fedkevych}
\email{ofedkevych@gsu.edu}
\affiliation{Physics and Astronomy Department, Georgia State University, Atlanta, GA 30303, USA.} 
\affiliation{Center for Frontiers in Nuclear Science, Stony Brook University, Stony Brook, NY 11794, USA.}
\affiliation{Jefferson Lab, Newport News, Virginia 23606, USA}

\author{Roli~Esha}
\email{roli.esha@stonybrook.edu}
\affiliation{Department of Physics and Astronomy, Stony Brook University, Stony Brook, NY 11794, USA.}
\affiliation{Center for Frontiers in Nuclear Science, Stony Brook University, Stony Brook, NY 11794, USA.}

\begin{abstract}
Jet substructure studies at the Large Hadron Collider have been used to constrain parton distribution functions, test perturbative QCD, measure the strong-coupling constant, and probe the properties of the quark-gluon plasma. We extend these studies to lower energies at the Relativistic Heavy Ion Collider that would additionally allow us to test existing models of non-perturbative physics. In this study, we present a PYTHIA8-based Monte Carlo study of substructure of jets produced in $p+p$ collisions at 200 GeV. The selection criteria are adapted for a feasible measurement at sPHENIX. We consider different types of jet substructure observables such as jet angularities and primary Lund Plane with a special focus on suppression of collinear radiation around emitting heavy quarks known as a dead cone effect. 
\end{abstract}
\maketitle

\section{Introduction}
Jets, described as collimated sprays of particles, encode rich information about hard QCD processes occurring at spatial scales significantly smaller than the proton radius. The internal structure of jets, often referred to as jet substructure, provides a unique probe into the fundamental properties of strong interactions. Over the past decade, the field of jet substructure has matured, yielding a wealth of applications. For a review, see~\cite{Marzani:2019hun, Apolinario:2024equ}.

Jet substructure observables have become important tools in a variety of contexts. They are instrumental in precision determination of the strong coupling constant~\cite{Britzger:2017maj, CMS:2013vbb, ATLAS:2017qir, ATLAS:2015yaa, CMS:2014mna}, provide constraints on parton distribution functions (PDFs)~\cite{ATLAS:2021qnl, ATLAS:2013pbc, CMS:2014qtp, CMS:2016lna, AbdulKhalek:2020jut, Harland-Lang:2017ytb, Pumplin:2009nk, Watt:2013oha}, and enable stringent tests of high-precision perturbative QCD calculations~\cite{CMS:2021iwu, ALICE:2021njq}. Moreover, they play a central role in searches for heavy hadronically decaying resonances predicted by theories beyond the Standard Model~\cite{Soper:2010xk, Godbole:2014cfa, Chen:2014dma, Adams:2015hiv}. In addition, jet substructure has been widely adopted as input for machine learning algorithms~\cite{Larkoski:2017jix, Kasieczka:2019dbj, Benato:2020sbi} and is extensively employed for particle tagging~\cite{Fedkevych:2022mid, Caletti:2021ysv, Dreyer:2021hhr, Cavallini:2021vot, Khosa:2021cyk, Dreyer:2020brq, Baron:2023hkp, Larkoski:2024uoc, Larkoski:2013eya, Gras:2017jty}. Finally, since the formation and evolution of jets are sensitive to the interactions between partons produced via hard scattering process with the dense medium formed in proton-nucleus ($p+\rm A$) or nucleus-nucleus ($\rm A+A$) collisions~\cite{Lapidus:2017dek,Zapp:2017ria,Tywoniuk:2017dzi,Casalderrey-Solana:2017mjg, Casalderrey-Solana:2019ubu, Mangano:2017plv,Qin:2017roz,Milhano:2017nzm,Chang:2017gkt,KunnawalkamElayavalli:2017hxo, Caucal:2019uvr, Caucal:2020uic,Caucal:2021cfb, Cunqueiro:2023vxl, Chien:2024uax, Brewer:2020och}, jet substructure has become an important tool for studies of the properties of Quark Gluon Plasma (QGP).

While jet substructure at high center-of-mass energies has been extensively studied at the LHC~\cite{ALICE:2017nij, ALICE:2018dxf, CMS:2011hzb, CMS:2013lhm, CMS:2013kfv, CMS:2013lua, CMS:2014qvs, CMS:2016php, CMS:2017eqd, CMS:2017pcy, CMS:2018mqn, CMS:2018fof, CMS:2018vzn, CMS:2018ypj, CMS:2019fak, ATLAS:2011kzm, ATLAS:2012nnf, ATLAS:2012am, ATLAS:2014lzu, ATLAS:2017nre, ATLAS:2017zda, ATLAS:2017xqp, ATLAS:2018jsv, ATLAS:2019dty, ATLAS:2019mgf}, similar investigations at the Relativistic Heavy Ion Collider (RHIC) remain relatively sparse~\cite{Kauder:2017mhg, STAR:2020ejj, STAR:2021lvw, Connors:2017ptx}. Consequently, jet substructure studies at RHIC offer a crucial platform for testing and refining existing models of non-perturbative QCD as highlighted in~\cite{Chien:2024uax} and would allow to test different regimes of QCD, in particular to separate contributions from hadronization and the soft background due to the underlying event. 

Two observables crucial to the measurement of jet substructure are the jet angularities~\cite{Berger:2003iw,Almeida:2008yp,Larkoski:2014pca} and the Primary Lund Plane (pLP) projections~\cite{Andersson:1988gp, Dreyer:2018nbf}. Jet angularities are a family of observables characterized by the exponent $\alpha$ given by $\lambda_\alpha = \sum_{i} z_{i} \left(\Delta_i / R\right)^{\alpha}$ where $z_i$ is the fraction of the jet momentum carried by its  $i-$th constituent, $\Delta_i$ is the angular distance from the jet axis for the $i-$th component and $R$ is the jet radius used to set the jet clustering algorithm. While $\alpha < 1$ are generally more sensitive to soft QCD effects, such as hadronization or underlying event contributions, $\alpha \geq 1$ tend to be less affected by non-perturbative contributions and hence are better suited to test analytical results obtained for partonic degrees of freedom, see \textit{e.g.} Refs.~\cite{Reichelt:2021svh, Caletti:2021oor, Chien:2024uax}.
 
On the other hand, unlike jet angularities, the pLP maps emissions within a jet onto a two-dimensional plane, providing insights into the dynamics of parton showers and QCD radiation. It encodes the relationship between the relative momentum scales and the collinearity of QCD emission. It allows for a separate study of hard wide-angle, soft wide-angle, soft-collinear and hard-collinear emissions inside of the jets and, as a consequence, to test different QCD regimes.

A major scientific motivation for the sPHENIX experiment is its broad program of reconstructed-jet physics measurements~\cite{Aidala:2012nz}. The large acceptance, hermetic hadronic calorimetry, high-efficiency tracking, and large data-taking rate of the detector will result in an enormous data sample of high-$p_T$ reconstructed jets. These high-statistics measurements will enable precise studies of dijet asymmetry, jet spectra, and jet substructure within the quark-gluon plasma (QGP), providing critical insights into parton energy loss mechanisms. The sPHENIX results will complement the extensive jet substructure analyses performed at higher energies at the Large Hadron Collider (LHC)~\cite{ALICE:2017nij,CMS:2018vzn,ATLAS:2018jsv}, enhancing our understanding of QGP properties across different energy regimes.

Since production of particles in a QCD cascade is predominantly given by soft and collinear QCD emissions, a fundamental property of QCD radiation generally known as a dead cone effect~\cite{Dokshitzer:1991fd,Dokshitzer:1995ev}, \textit{i.e.} the suppression of QCD radiation around massive quarks must change the QCD radiation pattern and, therefore, manifest itself in substructure of produced jets. However, despite an extensive theoretical study of the dead cone effect and jet substructure of heavy jets in general,  see \textit{e.g.}  Refs.~\cite{Lee:2019lge, Vaidya:2020lih, Cunqueiro:2022svx, Craft:2022kdo, Apolinario:2022vzg, Andres:2023ymw, Caletti:2023spr, Ghira:2023bxr, Gaggero:2022hmv, Wang:2023eer, Zhang:2023jpe, Dhani:2024gtx, Jiang:2024qno, Dainese:2024wix, Aglietti:2024zhg}, only a handful of measurements of jet substructure observables for jets seeded by heavy quarks is available up to date~\cite{CMS:2024kzm, ALICE:2022phr, CMS:2024gds}. In this paper, we address the potential of employing jet substructure studies, specifically the pLP projection, with the sPHENIX experiment at RHIC to study the dead cone effect,    
the first direct observation of which was recently reported by the ALICE collaboration~\cite{ALICE:2021aqk}. In the context of $b$ physics,~\cite{Cunqueiro:2018jbh} first proposed to use the pLP to look for the dead-cone effect.\footnote{For indirect observations of the dead-cone effect, see also refs.~\cite{DELPHI:1992pnf, OPAL:1994cct,  OPAL:1995rqo, SLD:1999cuj, DELPHI:2000edu, ALEPH:2001pfo, ATLAS:2013uet}. Also see the preliminary CMS measurements~\cite{CMS:2024kzm}.}.

% ------- Fig 1 -------
\begin{figure*}
\begin{minipage}{0.99\linewidth}
\includegraphics[width=0.9\textwidth]{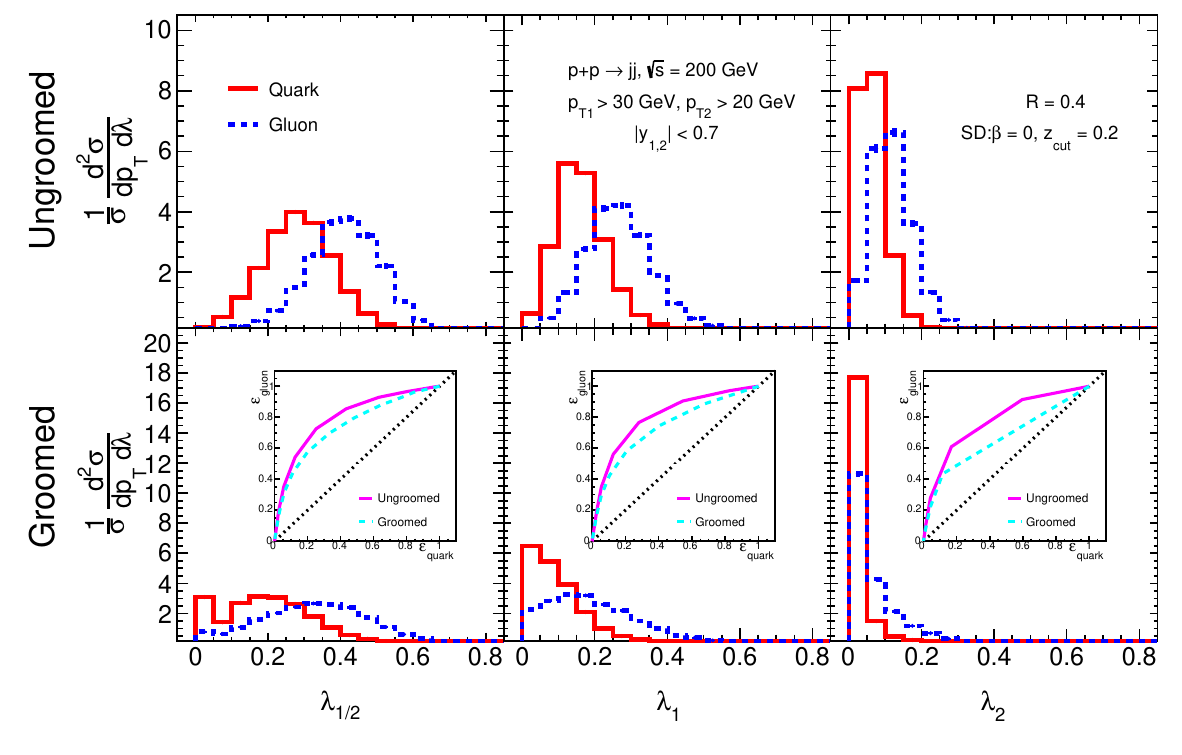}
\caption{Comparison of different angularities for ungroomed (top panel) and groomed (bottom panel) jets originating from quark and gluon. The inset shows the respective ROC curves for groomed and ungroomed jets.}
\label{fig:qg_ang}
\vspace{1cm}
%-------- Fig. 2 --------
\includegraphics[width=0.9\textwidth]{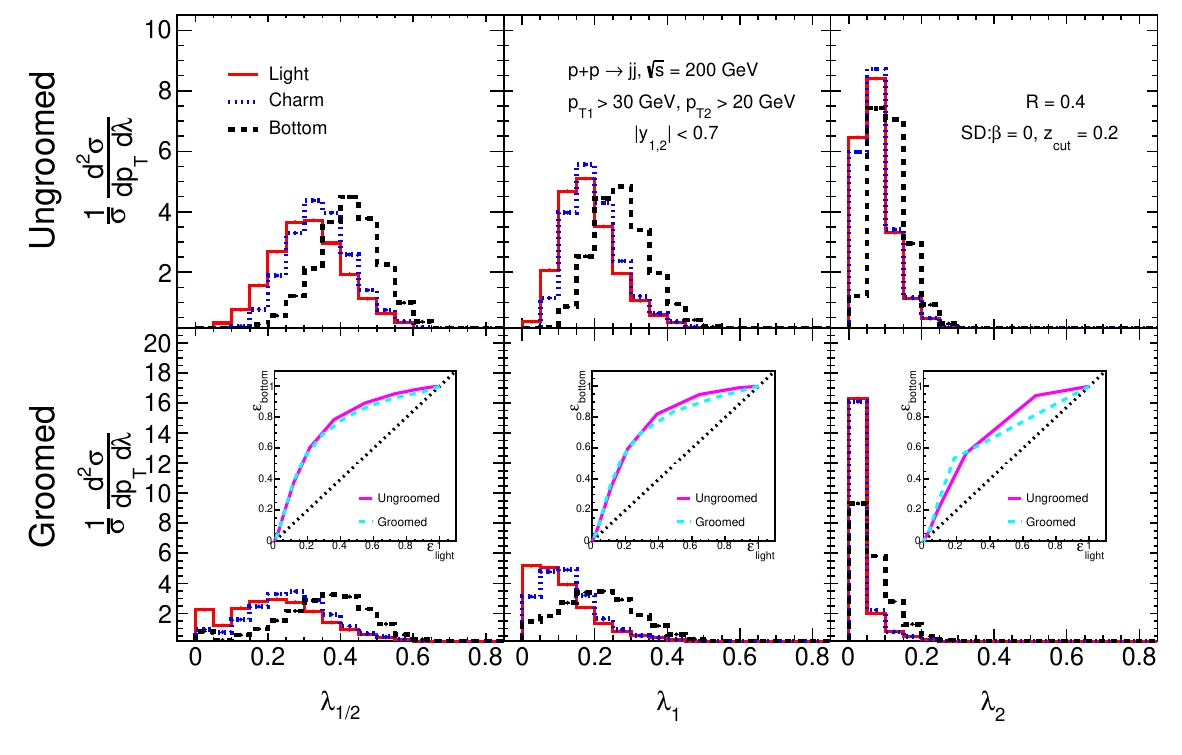}
\caption{Comparison of different angularities for ungroomed (top panel) and groomed (bottom panel) jets originating from light, charm, and bottom quarks. The inset shows the respective ROC curves for groomed and ungroomed jets.}
\label{fig:lcb_ang}
\end{minipage}
\end{figure*}

\section{Monte Carlo setup and event selection}
In this study, jets seeded by massive $b$- and $c$-quarks are compared against jets initiated by light flavors($u$, $d$, $s$). In addition, quark jets are also compared with gluon jets. The samples of pseudo data using \mbox{\pythia8}~\cite{Sjostrand:2014zea, Bierlich:2022pfr} Monte Carlo event generator with a Detroit tune~\cite{Aguilar:2021sfa} is generated for $p+p$ collisions at 200 GeV. The selection of PDF is set to 17, which corresponds to NNPDF3.1 LO set~\cite{NNPDF:2017mvq} and the configuration of multipartonic interactions is characterized by a reference transverse momentum scale of 1.40 GeV, a core radius of 0.56, and a core fraction of 0.78. The \texttt{PhaseSpace:pTHatMin = 20} parameter is implemented to impose a minimum transverse momentum scale of 20~GeV, thereby serving as a low-$p_T$ cutoff that ensures only sufficiently energetic partonic interactions are included in the event generation. Color reconnection effects are modeled to accurately represent the underlying event dynamics.

Three different samples of $b$-jets, $c$-jets and light-jets with jets produced via $2\rightarrow2$ LO QCD di-jet production processes are simulated where each di-jet pair is required to satisfy the following fiducial cuts: $p_{T1} > 30$~GeV, $p_{T2} > 20$~GeV, and $|\eta_{\rm jet}| < 0.7$ which is in agreement with sPHENIX acceptance~\cite{Okawa:2023asr}. The transverse momentum difference between leading and subleading jets is inspired by currently available theoretical studies~\cite{Reichelt:2021svh, Chien:2024uax}\footnote{The transverse momentum imbalance cuts as in Refs.~\cite{Reichelt:2021svh, Chien:2024uax} were used to ensure to avoid large K-factors and associated numerical instabilities.}. Jet formation is modeled by including initial and final state radiation, multiple parton interactions, and hadronization processes within the simulation. The final-state particles are clustered into jets using  \mbox{anti-$k_t$}  algorithm~\cite{Cacciari:2008gp} with jet radius parameter $R = 0.4$ and standard \mbox{$E$-scheme} recombination using the \fastjet~\cite{Cacciari:2011ma} library.
Two versions of each jet are considered; namely ungroomed (taken as it is), and groomed, where the \softdrop algorithm with parameters $\zc = 0.2$ and
$\beta = 0$~\cite{Marzani:2017mva, Larkoski:2014wba} are used with the Cambridge--Aachen (C/A) algorithm \cite{Dokshitzer:1997in, Wobisch:1998wt}
for jet reclustering as implemented in the \fjcontrib~package as the groomer. The \softdrop\ algorithm is an iterative procedure that consists of two steps:
\begin{enumerate}
    \item Recluster the jet into two subjects.
    \item Check the \softdrop\ condition:
    \begin{eqnarray}
        \frac{\min(p_{T1},\, p_{T2})}{p_{T1} + p_{T2}} > z_{\text{cut}} \left( \frac{\Delta_{12}}{R} \right)^\beta,
    \end{eqnarray}
    where $\Delta_{12}$ is a standard (pseudo)rapidity-azimuth distance between two branches and $R$ is a jet radius.
    If the soft branch carries a fraction of jet $p_T$ smaller than $z_{\text{cut}} \left( \frac{\Delta_{12}}{R} \right)^\beta$, discard it and repeat the reclustering procedure. 
   	If the \softdrop\ condition is met, end the procedure.
\end{enumerate}

The \softdrop procedure is well defined from a theoretical point of view as it satisfies the concept of Infrared and Collinear (IRC) safety and is known to reduce the impact of soft wide-angle radiation on jet substructure which, in turn, makes it less sensitive to the non-perturbative effects such as hadronization and underlying event contribution~\cite{Larkoski:2014wba, Marzani:2017mva}.

For each event, jet substructure observables are calculated for both the leading and subleading jets. Subsequently, jets are categorized as light quark, charm quark, bottom quark, or gluon jets by matching each jet to initial partons from the hard scattering in the \textsc{Pythia} event record, provided that the angular distance, $\Delta$ 
between the jet axis and hard parton seeding the jet is smaller than 0.4. 

\section{Jet Angularities}
Jet angularities are defined as 
\begin{equation}\label{eq:ang-def}
\la= \sum_{i \in \text{jet}}\left(\frac{p_{\rm T,i}}{\sum_{j \in \rm jet} p_{\rm T,j}}\right)\left(\frac{\Delta_i}{R} \right)^\alpha\,,
\end{equation}
where the sum runs over all jet constituents, $R$ is the jet radius and 
\begin{equation}\label{eq:dist-def}
\Delta_i=\sqrt{(y_i-y_\text{jet})^2+(\phi_i-\phi_\text{jet})^2}\, 
\end{equation}
is the Euclidean azimuth-rapidity distance of particle $i$ from the jet axis.  The requirement of  IRC safety implies $\kappa = 1$ and $\alpha > 0$. Therefore, three commonly used cases are considered namely, $\lambda_{1/2}$ (LHA), $\lambda_1$ (Jet Width), and $\lambda_2$ (Jet Thrust) \cite{Larkoski:2014pca, Andersen:2016qtm}, thereby controlling the sensitivity to the distribution of particles within the jet. For the reference axis, the winner-takes-all (WTA) recombination scheme~\cite{Larkoski:2014wba, Sjostrand:2004ef} is employed when $\alpha = 2$. 

The comparison of angularities for quark and gluon jets is shown in Figure~\ref{fig:qg_ang}. The top panel shows the distributions for $\lambda_{1/2}$, $\lambda_1$, and $\lambda_2$ for ungroomed jets, while the bottom panel shows the same for groomed jets. The quark-initiated jets are represented by solid red lines, whereas gluon-initiated jets are given by blue dashed ones. The peak position of distributions for gluon jets tends to be shifted toward larger values of jet angularities as compared to the quark jets. This behavior is consistent with previous results available in the literature~\cite{Larkoski:2013eya, Gras:2017jty, Marzani:2019hun}, and is caused by different emission patterns of quark and gluons. In order to quantitatively estimate the efficiency of quark-gluon separation for the RHIC setup, the Receiver Operating Characteristic (ROC) curves are presented as insert plots in Figure~\ref{fig:qg_ang}. Each point on a ROC curve represents a fraction of the signal (in our case gluon) and background (in our case quark) jets selected after triggering on jets with angularity value $\lambda \geq \lambda_{\rm cut}$, where $\lambda_{\rm cut}$ is varied in between 0 and 1. The ROC curves for ungroomed jets are presented by solid magenta lines, whereas ROC curves corresponding to \softdrop~jets are given by dashed cyan lines. The best operational  point corresponds to 
$\lambda_{\rm cut} = 0.3$, $\epsilon_q = 0.43$, $\epsilon_g = 0.85$. The possibility of efficient separation between quark and gluon jets is suggested by the simulation at RHIC energies which, in turn, would allow for testing of existing approaches to perturbative and non-perturbative descriptions of jets produced at collision energies much lower compared to the LHC.

The case for groomed jets is shown in the bottom panel of Figure~\ref{fig:qg_ang}. As seen, in general, the performance of quark-gluon discrimination for groomed jets is somewhat smaller compared to the ungroomed ones. This is in agreement with the expectations, since \softdrop~grooming reduces the fraction of soft emissions inside the jet, which partially contributes to the difference between quark and gluon jets (quark jets tend to be more collimated, whereas gluon jets have a larger particle spread).  It should also be noted that the quark distributions after grooming demonstrate a significant change in their shape: namely, the groomed distributions have peaks around the first bin. This effect is much stronger for the quark jets which can be explained by the fact that quark jets, in general, tend to have fewer constituents comparing to the gluon jets. Moreover, at RHIC beam energies, quark jets have a much smaller amount of energy available for radiation compared to the LHC. As a consequence, it is more likely to create a quark jet made of a single hard particle and a few soft particles which are eliminated after \softdrop application. This would lead to quark jets with a single constituent and, therefore, zero angularity value\footnote{Additionally, as it was pointed out in Ref.~\cite{Chien:2024uax}, the low center-of-mass collision energy at RHIC implies significant suppression of uniform soft radiation from beam remnants which, in turn, contributes to the peak creation at low $\lambda$ values.}.

The presence of non-negligible quark masses for $c$- and $b$-jets leads to non-trivial modification of the QCD emission cascade~\cite{Dokshitzer:1991fd, Dokshitzer:1995ev, Ellis:1996mzs, Catani:2000ef, Catani:2002hc, Dhani:2023uxu, Craft:2023aew} which, in turn, changes the shape of the jet substructure observables~\cite{Lee:2019lge, Maltoni:2016ays, Llorente:2014bha, Li:2017wwc, Li:2021gjw, Craft:2022kdo, Cunqueiro:2022svx, Caletti:2023spr, Blok:2023ugf, Zhang:2023jpe,  Dhani:2024gtx, Aglietti:2024zhg, Wang:2024yag}. \pythia takes into account modifications of the parton shower due to the presence of non-negligible quarks masses~\cite{Norrbin:2000uu, Bierlich:2022pfr}. Even though, as it was argued in~\cite{Bierlich:2022pfr}, the \pythia approach to radiation of heavy quarks~\cite{Norrbin:2000uu} does not fully respect the dead-cone effect, it can still be used to obtain a quantitative estimate of jet substructure of heavy jets~\cite{ALICE:2021aqk, Dhani:2024gtx}. Therefore, these MC simulations for jet substructure of heavy jets are a good starting point for future jet substructure studies at RHIC. More detailed study would require comparison against other existing MC models and theoretical calculations~\cite{Gieseke:2003rz, Hoang:2018zrp, Cormier:2018tog, Schumann:2007mg, Lee:2019lge, Dhani:2024gtx}.

The comparison of angularities for jets seeded by light ($u$, $d$, $s$), charm, and bottom quarks is shown in Figure~\ref{fig:lcb_ang}. The distributions for $\lambda_{1/2}$, 
$\lambda_1$ and $\lambda_2$ for ungroomed jets are given in the top panel, while the bottom panel shows the same for groomed jets. The results for $b$-jets tend to shift towards larger angularity values as compared to the light-jets, thereby achieving discrimination efficiency similar to those of angularity-based quark-gluon taggers in Figure~\ref{fig:qg_ang}. The observed shift of $b$-jet distributions is caused by two effects of perturbative and non-perturbative nature, correspondingly. The perturbative effect is a suppression of collinear radiation around the massive quarks for observable values smaller or equal to the dead-cone endpoint $m_Q / p_{T, \rm{jet}} R$ which implies that the contribution of parton shower to the angularity values with $\lambda \leq m_Q / p_{T, \rm{jet}} R$ is suppressed~\cite{Dhani:2024gtx}. A second effect is related to decays of produced $B$-hadrons: at RHIC energies, in the case of $b$-jets the probability of producing a jet made of a single $b$-quark is non-negligible. Such parton-level jets after hadronization will often convert into jets containing a single $B$-hadron. In the case of stable $B$-hadrons, the aforementioned events would contribute to a very first angularity bin (events with  $\lambda = 0$). However, because $B$-hadrons are unstable their decay products give rise to non-trivial modification of the jet substructure leading to transfer of events from the first angularity bin towards larger angularity bins, contributing to the peak of the distribution~\cite{Vaidya:2020lih, Dhani:2024gtx}. Therefore, as it was argued first in~\cite{Vaidya:2020lih}, the question of $B$-hadron decays needs to be carefully addressed by both theorists and experimentalists. In this study, perspectives of jet substructure studies for the sPHENIX experiment are addressed and a detailed discussion on the role of $B$-hadron decays is refrained and left to future investigations.

Finally, it should be noted that, in Figure~\ref{fig:lcb_ang}, the ROC curves obtained for groomed distributions mostly tend to be very similar to the ones obtained for the ungroomed jets. This behavior of the ROC curves is different from the results presented in Figure~\ref{fig:qg_ang}. However, it can be explained by the fact that both the sources of the differences between heavy and light jets (suppression of radiation and $B$-hadron decays) affect the collinear region which is not sensitive to \softdrop grooming, whereas in case of quark-gluon tagging, a significant portion of difference between quark and gluon jets comes from soft emissions which are subtracted by the grooming procedure.

% ------- Fig 3 -------
\begin{figure*}
\begin{minipage}{0.99\linewidth}
\includegraphics[width=0.7\linewidth]{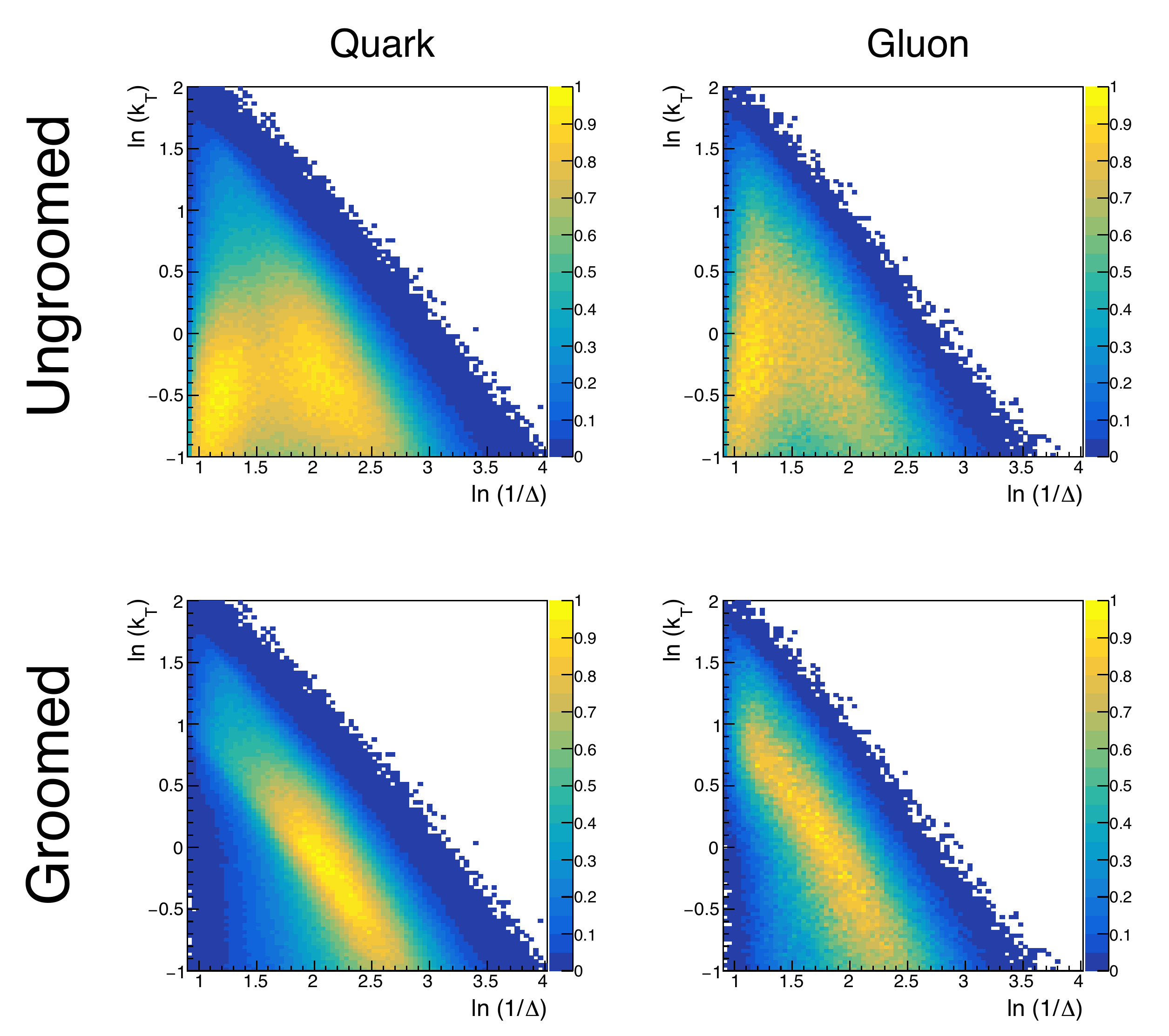}
\caption{Primary Lund Plane projections for ungroomed (top panel) and groomed (bottom panel) jets originating from quarks (left), and gluons (right).}
\label{fig:lp_qg}
\vspace{1cm}
%-------- Fig. 4 --------
\includegraphics[width=0.95\linewidth]{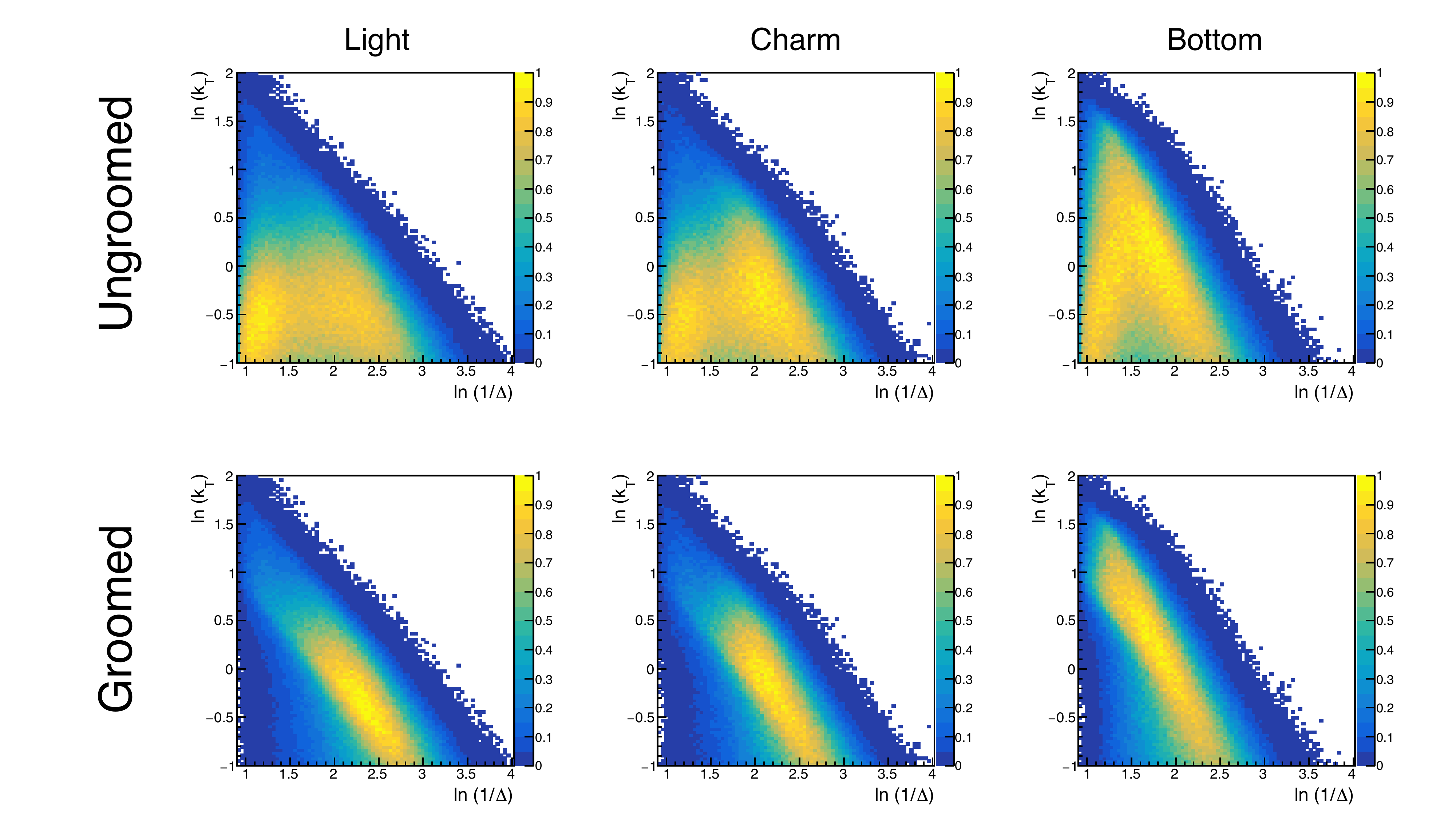}
\caption{Primary Lund Plane projections for ungroomed (top panel) and groomed (bottom panel) jets originating from light quarks (left), charm quarks (center), and bottom quarks (right).}
\label{fig:lp_lcb}
\end{minipage}
\end{figure*}

\section{Primary Lund Plane}
The pLP is a comprehensive tool that captures the kinematics and dynamics of QCD emissions within jets. It is made with axes that represent the logarithm of the transverse momentum fraction, encoding the relative momentum scale of emissions, and the logarithm of the inverse emission angle, representing the angular separation of emissions within the jet. Each emission within a jet is represented as a point on the plane, where its coordinates are determined by its transverse momentum and angular separation relative to the jet axis.

The pLP is constructed by reclustering the selected jet using the Cambridge-Aachen (C/A) jet algorithm. The declustering history of the hardest branch is followed and recorded at each splitting: 
\begin{eqnarray}
	\Delta_{ab} &\equiv& \sqrt{\left(y_a - y_b\right)^2 + \left(\phi_a - \phi_b\right)^2},\\
	k_T &\equiv& p_{\text{T}b} \, \Delta_{ab},
\end{eqnarray}
where $y_{a,b}$, $\phi_{a,b}$ are rapidity and azimuthal angle of subjects $a$ and $b$ correspondingly and $p_{T,a} > p_{T,b}$ This procedure is performed for both leading and subleading jet in our sample.

The pLP distribution of quark and gluon jets is shown in Figure~\ref{fig:lp_qg}. The ungroomed distributions for quarks (left) and gluons (right) are compared in the top panel, while the respective groomed distributions are shown in the bottom panel. Span through the $x-$axis, it is seen that ungroomed quark jets are more collinear ($\ln (1/\Delta) > 2$) as compared to ungroomed gluon jets. This is very representative of the difference in their radiation patterns. For the groomed pLP, the difference tends to become milder as the soft emissions are eliminated by \softdrop.

The dead cone effect is manifested as a suppression of gluon radiation in the vicinity of massive quarks ~\cite{Dokshitzer:1991fd, Dokshitzer:1995ev, Ellis:1996mzs}. This suppression arises because the quark mass introduces an effective angular cutoff, limiting gluon emissions at small angles relative to the trajectory of the heavy quark. In Figure~\ref{fig:lp_lcb}, pLP distributions for jets initiated by light quarks (left), charm quarks (middle), and bottom (right) quarks, the manifestation of the dead cone effect is shown for ungroomed (top panel) and groomed (bottom panel) jets. Specifically, a significant suppression of collinear radiation within the logarithmic range of $\ln(1/\Delta) \approx 2.5$ to $3$ for $b$-jets is revealed, a feature markedly less pronounced in jets seeded by lighter quarks. Additionally, this suppression is notably evident at low values of $\ln(k_T)$. The concurrent suppression in both $\ln(1/\Delta)$ and $\ln(k_T)$ underscores the mass-dependent modification of the radiation pattern inherent to heavy quarks. 

\section{Conclusion}
In this study, we performed a detailed \pythia-based Monte Carlo analysis of heavy-flavor jet substructure for $p+p$ collisions at 200 GeV, leveraging the capabilities of the sPHENIX detector. Focusing on jet angularities and the primary Lund Plane (pLP) projection, we have successfully identified the suppression of collinear radiation around heavy quarks, a manifestation of the dead cone effect. Our results reveal distinct differences in the radiation patterns of $b$-jets compared to light and charm jets and between quark and gluon jets, aligning with theoretical predictions and recent experimental observations. In addition to providing a framework for the measurement of jet substructure, we were able to establish the feasibility of the experimental observation of the dead cone effect at sPHENIX. These predictions can also be used for QCD-inspired observables for the study of tagged-jets at RHIC energies.

\section*{Acknowledgments}
We thank Axel Drees and Yang-Ting Chien for very useful discussions. RE and ZY acknowledge the support from US Department of Energy Contract No.~DE-FG02-96ER40988. The work of OF is supported in part by the US Department of Energy Contract No.~DE-AC05-06OR23177, under which Jefferson Science Associates, LLC operates Jefferson Lab, and by the Department of Energy Early Career Award grant DE-SC0023304.

\bibliographystyle{apsrev4-2}
\bibliography{references}
\end{document}